\documentstyle[pre,aps,twocolumn,epsf]{revtex}

\def\LM#1#2{\left|\begin{array}{l}{#1}\\[1ex]{#2}\end{array}\right.}

\begin{document}
\draft
\title{Subdiffusion-limited reactions}
\author{S. B. Yuste\cite{santos} and Katja Lindenberg}
\address{Department of Chemistry and Biochemistry 0340, 
University of California San Diego, La Jolla, California 92093-0340}
\date{\today}

\maketitle
\begin{abstract}
We  consider the coagulation dynamics $A+A\rightarrow A$ and $A+A
\rightleftharpoons A$ and the annihilation dynamics $A+A \rightarrow 0$
for particles moving subdiffusively in one dimension.  This scenario
combines the ``anomalous kinetics" and ``anomalous diffusion" problems,
each of which leads to interesting dynamics separately and to even more
interesting dynamics in combination. Our analysis is based on the
fractional diffusion equation.
\end{abstract}
\pacs{PACS numbers: 02.50.Ey,82.40.-g,05.60.-k,05.70.Ln}

Diffusion-limited reactions in constrained geometries 
have been studied intensely because they exhibit ``anomalous kinetics,"
that is, behavior different from
that predicted by the laws of mass action in well-stirred
systems~\cite{book}.
Among the simplest and most extensively studied are
single species diffusion-limited coagulation ($A+A\rightarrow A$ or $A+A
\rightleftharpoons A$)~\cite{DBA} and annihilation ($A+A \rightarrow
0$)~\cite{DBA,ours}.  These
reactions, which show anomalous behavior in one dimension,
are of particular theoretical interest because they
lend themselves to {\em exact} solution in one
dimension~\cite{DBA,TorMcCJPC,Spouge,MasAvrPLA,wenew}.
The anomalies are typically displayed in two ways:
one is through the time dependence of the reactant
concentration $c(t)$, which for the
$A+A\rightarrow A$ and the $A+A\rightarrow 0$ reactions decays as
$t^{-1/2}$ in one dimension instead of the law of mass action decay
$t^{-1}$.  The other is through the interparticle distribution function
$p(x,t)$, which is the (conditional) probability density for
finding the nearest particle at a distance $x$ on one side of a given
particle.  This function scales as $x/t^{1/2}$, in typical diffusive
fashion.  In one dimension a gap develops around each particle 
that leads to a more ordered spatial distribution than the
exponential distribution implicit in well-stirred systems and ``explains"
the relative slowing down of the reaction.

In a parallel development, the problem of ``anomalous diffusion" has also
attracted a great deal of
attention~\cite{Kehr,HilferEd,MetKlaPhysRep}. 
The universally accepted characterization of
anomalous (as in ``not ordinary") diffusion
is through
the mean squared displacement of a process $x(t)$ for large $t$,
\begin{equation}
\left< x^2(t)\right> \sim \frac{2K_\alpha}{\Gamma(1+\alpha)} t^\alpha .
\label{meansquaredispl}
\end{equation}
Ordinary diffusion ($\alpha=1$, $K_1\equiv D$) follows Gaussian
statistics and Fick's second law leading to linear growth of
$\left< x^2(t)\right>$ with time.  Anomalous diffusion is characterized
by a nonlinear dependence.
If $0<\alpha<1$ the process is subdiffusive or
dispersive; if
$\alpha>1$ it is superdiffusive.  Anomalous diffusion is associated with
many physical systems and is not due to any single universal cause, but
it is certainly ubiquitous.   Nor is anomalous diffusion modeled in a
universal way; among the more successful approaches to the subdiffusive
problem have been continuous time random walks with non-Poissonian
waiting time distributions~\cite{Kehr}, and fractional
dynamics approaches in which the diffusion equation is replaced by
a generalized diffusion equation~\cite{HilferEd,MetKlaPhysRep}.  
Some connections between these two approaches have recently been
clarified~\cite{BarMetKlaPRE}.

In this work we consider a combination of these two phenomena, namely,
the (one-dimensional) kinetics of $A+A$ reactions of particles that
move subdiffusively.  We pose two questions: (1) How does the
reactant concentration evolve in time? (2) How does the interparticle
distribution function evolve in space and time? Some aspects of
this problem have been considered previously using the waiting time
distribution approach~\cite{Blumen,AlemanyJPA}.  The solutions require
approximations relating the reactant concentration to the distinct
number of sites visited by a particle~\cite{Blumen}, or the
waiting time distributions for single particles to the waiting time
distributions for relative motion~\cite{AlemanyJPA}.
Here we adapt the 
fractional dynamics approach to the problem and take advantage of the fact
that the resulting generalized diffusion equations can 
be solved in closed form.  We consider both
coagulation and annihilation reactions.

Consider first the coagulation reaction $A+A \rightarrow A$ when the
particles move by ordinary diffusion. The probability distribution
function for the position $y$ of any {\em one} $A$ particle {\em in
the absence of reaction} obeys the diffusion equation
\begin{equation}
\frac{\partial}{\partial t} P(y,t)= D \frac{\partial^2}{\partial y^2} P(y,t).
\label{Peqdifun}
\end{equation}
The coagulation problem can be formulated in terms of the probability 
$E(x,t)$ that an interval of length $x$ is empty of particles
at time $t$.  This ``empty interval" function for a diffusion-limited
reaction (i.e., one with immediate reaction upon encounter) obeys the 
diffusion equation~\cite{DBA}
\begin{equation}
\frac{\partial}{\partial t} E(x,t)= 2D \frac{\partial^2}{\partial x^2} E(x,t).
\label{Eeqdifun}
\end{equation}
The derivation of this equation is
straightforward and recognizes that an empty interval is shortened
or lengthened by movement of particles in and out at either end
according to the dynamics described by Eq.~(\ref{Peqdifun}). 
The empty interval dynamics is thus
essentially the same as that of individual particles in the absence of
reaction, but with a diffusion coefficient
$2D$ that reflects the fact that
the relative motion of two diffusive particles involves twice the
diffusion coefficient of each particle alone.  The coalescence reaction
implies the boundary condition $E(0,t)=1$, and $E(\infty, t) =0$ as long
as the concentration is nonvanishing.

From $E(x,t)$ one obtains the concentration of particles
\begin{equation}
c(t)=-\left. \frac{\partial E(x,t)}{\partial x}\right|_{x=0} ,
\label{cE}
\end{equation}
and the interparticle distribution function
\begin{equation}
p(x,t)= \frac{1}{c(t)} \frac{\partial^2 E(x,t)}{\partial x^2}.
\label{pE}
\end{equation}
Equation~(\ref{Eeqdifun}) and the boundary conditions can readily be
generalized in a number of ways~\cite{multi3}, in particular 
to reversible coagulation (A+A$\rightleftharpoons $A)
and nucleation~\cite{DoeAvrPRL} and even to processes with three-site
interactions~\cite{Henkel}.

The motion of a subdiffusive particle (in
the absence of reaction) is described by the fractional diffusion
equation~\cite{DoeAvrPRL,Balakrishnan,SchWysJMP,MetBarKlaPRL,MetBarKlaEPL}
\begin{equation}
\frac{\partial }{\partial t} P(y,t)=
~_{0}\,D_{t}^{1-\alpha } K_\alpha \frac{\partial^2}{\partial y^2} P(y,t)
\label{Pfracdifu}
\end{equation}
where  $~_{0}\,D_{t}^{1-\alpha } $ is the Riemann-Liouville operator:
\begin{equation}
~_{0}\,D_{t}^{1-\alpha } P(y,t)=\frac{1}{\Gamma(\alpha)}
\frac{\partial}{\partial t} \int_0^t d\tau
\frac{P(y,\tau)}{(t-\tau)^{1-\alpha}}
\end{equation}
and $K_\alpha$ is the generalized diffusion coefficient 
that appears in Eq.~(\ref{meansquaredispl}).
Some limitations of this description have been discussed
recently~\cite{BarMetKlaPRE}.

The construction of the kinetic equation for $E(x,t)$ for subdiffusive
particles proceeds along arguments analogous to those used in
the diffusive case.  Again, one follows the motion of the particles in
and out of the ends of the empty interval according to the dynamics
(\ref{Pfracdifu}).  This readily leads to the fractional diffusion
evolution equation for the empty intervals for subdiffusion-limited reactions
\begin{equation}
\frac{\partial }{\partial t} E(x,t)=
~_{0}\,D_{t}^{1-\alpha } 2K_\alpha \frac{\partial^2}{\partial x^2} E(x,t).
\label{Ebasica}
\end{equation}

The solution of Eq.~(\ref{Ebasica}) with the boundary conditions 
$E(0,t)=1$, and $E(\infty, t) =0$ 
can be expressed in terms of the Fox H-functions~\cite{SchWysJMP}. 
In Laplace transform space (indicated by a tilde over the function) the
solution is 
\begin{eqnarray}
\widetilde{E}(x,u)&=&\frac{s}{2 u}
\int_0^\infty dy \left[ e^{-|x-y| s }  -  e^{-|x+y|s } \right]   E_0(y)  
\nonumber \\ [6pt]
&&+ \frac{1}{u}   \exp\left[-x s \right]
\label{EsolLap}
\end{eqnarray}
where $s\equiv u^{\alpha/2}/\sqrt{2K_\alpha}$ and  $E_0(x)\equiv E(x,0)$.
From Eqs.~(\ref{cE}) and  (\ref{EsolLap}) one finds:
\begin{equation}
\widetilde c(u)=\frac{\lambda}{u} \left[1-\widetilde
p_0\left(\frac{u^{\alpha/2}}{\sqrt{2K_\alpha}}\right)\right]
\label{cu}
\end{equation}
where $\lambda\equiv c(0)$ and $\widetilde p_0(s)$ is the {\em spatial}
Laplace transform of the initial interparticle distribution function
$p_0(x)=p(x,0)$.

A commonly considered initial interparticle distribution is the random
(Poisson) distribution of average concentration $\lambda$, 
$p_0(x)=\lambda e^{-\lambda x}$.  For this initial distribution
$\widetilde c(u) =\lambda/\left(u+\lambda \sqrt{2K_\alpha}
u^{1-\alpha/2}\right)$ and $c(t)$ is given in closed form in terms of
the Mittag-Leffler function~\cite{MetKlaPhysRep,PodlubMainar} of
parameter $\alpha/2$:
\begin{equation}
c(t)= \lambda E_{\alpha/2}   \left(-\lambda \sqrt{2K_\alpha} t^{\alpha/2}
\right).
\label{ctML}
\end{equation}
When $\alpha=1$ one recovers the usual result for diffusion-limited
coagulation~\cite{TorMcCJPC,Spouge} since the
Mittag-Leffler function of parameter $1/2$ is
$E_{1/2}(-x)=\exp(x^2)\text{erfc}(x)$.

\begin{figure}
\begin{center}
\leavevmode
\epsffile{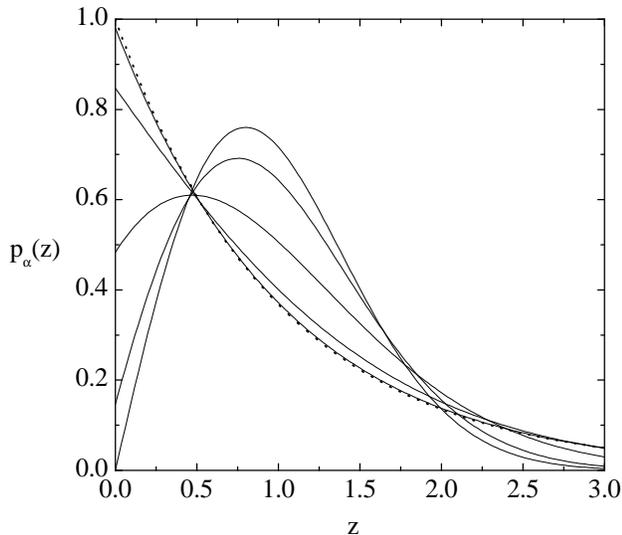}
\end{center}   
\caption{Long-time scaled interparticle distribution function
for several values of the anomalous diffusion exponent.  Proceeding
upward from lowest to highest curves along the $y$ axis intersection:
$\alpha=1, 0.95, 0.8, 0.5, 0.2$.  
Note that the distribution for $\alpha=0.2$ on this scale is nearly
indistinguishable from the completely random distribution $\exp(-z)$
(dotted curve).
} 
\label{fig:IPDF}
\end{figure}

The $u\rightarrow 0$ expansions of Eqs.~(\ref{EsolLap}) and (\ref{cu})
are readily seen to be independent of the initial distribution $p_0(x)$
and can be Laplace inverted to yield the asymptotic results for large $t$:
\begin{equation}
c(t)\sim \frac{t^{-\alpha/2}}{\sqrt{2K_\alpha}\Gamma
\left(1-\frac{\alpha}{2}\right)}
\label{ctasin}
\end{equation}
and
\begin{equation}
p(x,t)c(t)\sim  \frac{1}{x^2}
H^{10}_{11}\left[\frac{x}{\sqrt{2K_\alpha} t^{\alpha/2}}
	\LM{(1 ,\frac{\alpha}{2})}{(2,1)}   \right],
\label{Extasin}
\end{equation}
where $H$ is the Fox
H-function~\cite{HilferEd,MetKlaPhysRep,SchWysJMP,MathaiSaxena}.
Furthermore, with Eqs.~(\ref{ctasin}) and (\ref{Extasin}) 
we find
\begin{equation}
p_\alpha(z) \sim \Gamma^2\left(1-\frac{\alpha}{2}\right)
H^{10}_{11}\left[\Gamma(1-\frac{\alpha}{2})z
\LM{\left(1-\alpha ,\frac{\alpha}{2}\right)}{(0,1)}   \right]
\label{pztasin}
\end{equation}
where $z\equiv c(t)x$ is the scaled interparticle distance
and $p_\alpha(z)dz\equiv p(x,t)dx$.
This stationary form is shown in Fig.~\ref{fig:IPDF}
for  several values of the diffusion exponent $\alpha$.

The interparticle distribution function conveys the interesting
``anomalies" of the problem most clearly.  For a random distribution of
particles on a line this distribution is exponential.  In particular,
the most probable interparticle gaps are the smallest.
For diffusion-limited reactions on a line it is well known that 
the scaled distribution deviates
in two ways from the exponential behavior. First, a gap develops around
each particle, and the distribution vanishes near the origin (see
$\alpha=1$ curve in the figure), indicating an ``effective repulsion"
of particles.  Second, the probability of large gaps decays much more
rapidly than exponentially: the decay goes as a power of $\exp(-z^2/2)$.  
In the subdiffusive case decreasing $\alpha$ leads to the diminution 
of the gap around each particle, that is, to a weakening of the
effective repulsion and to a behavior that appears closer to that of
a random distribution in the short-interparticle-distance behavior.
This is evident in the progression of the curves with decreasing
$\alpha$ shown in the figure.  Furthermore, 
the probability of large gaps decays as a power of
$\exp(-x^{2/(2-\alpha)})$, thus neatly interpolating 
between the purely random exponential decay 
state as $\alpha\rightarrow 0$ (since then
$p(x,t)\rightarrow c(t) e^{-c(t) x}$)
and the more ordered state corresponding to diffusive particles at
$\alpha=1$.  Note that the congruence of the curves at $z=1/2$ is
{\em not} exact although the almost-congruence is certainly intriguing
(and may be
related to the special but physically unclear role played by the initial
concentration
$c_{eq}/2$ in the reversible coagulation problem that reaches an
equilibrium state at concentration $c_{eq}$, see~\cite{DBA}
and Eq.~(\ref{EreverGV}) {\em et seq.}).

The arguments presented so far are applicable to the irreversible
coagulation reaction $A+A\rightarrow A$.   The empty interval method 
can not be applied to the annihilation reaction $A+A\rightarrow 0$
because annihilation leads to a discontinuous growth of empty intervals.
However, recently a new method of odd/even intervals has been introduced 
that leads to exact solution in the diffusion-limited case.  It 
is based on the construction of an equation for $r(x,t)$,
the probability that an arbitrary interval of length
$x$ contains an even number of particles at time
$t$~\cite{MasAvrPLA,wenew}.
Again, because $r(x,t)$ changes only by the movement of particles in or
out of the ends of the interval, arguments similar to those that lead to
Eq.~(\ref{Eeqdifun}) lead to exactly the same equation for $r(x,t)$ but
with the boundary conditions $r(0,t)=1$ (as for $E(0,t)$) and
$r(\infty,t)$=1/2.  The concentration of particles is related to
$r(x,t)$ precisely as in Eq.~(\ref{cE}).  

The method of odd/even intervals can again be directly extended to the
subdiffusive problem, where $r(x,t)$ satisfies the same fractional
diffusion equation as $E(x,t)$ with appropriately modified
boundary conditions. Closed solution is again possible and, for a random
initial distribution with average initial concentration $\lambda$ leads to a
form slightly modified from the result~(\ref{ctML}):
\begin{equation}
c(t)= \lambda E_{\alpha/2}   \left(-2\lambda \sqrt{2K_\alpha} t^{\alpha/2}
\right).
\label{ctMLnew}
\end{equation}

It is appropriate to make contact with the ingenious work
of Spouge~\cite{Spouge}.  He introduced a single formalism 
to calculate the particle concentration $c(t)$ for both coagulation and
annihilation.  His approach is based on the
probability $a(x,t)$ that two particles, one starting at a distance $x$ 
from the other, have met by time $t$.  In ordinary diffusion this
quantity also obeys the diffusion equation with diffusion coefficient
$2D$ with appropriate boundary conditions.  Again, his method can
be generalized to anomalous diffusion and the resulting $a(x,t)$ 
is~\cite{MetKlaPhysA}
\begin{equation}
a(x,t)=H^{10}_{11}\left[\frac{x}{\sqrt{2K_\alpha} t^{\alpha/2}}
	\LM{(1 ,\alpha/2)}{(0,1)}   \right] . 
\label{survi}
\end{equation}
Spouge's prescription applied to this
subdiffussive situation leads, for an initially random distribution, to
the results (\ref{ctML}) and (\ref{ctMLnew}) for coagulation and
annihilation respectively.

The irreversible processes can be generalized in a variety of ways for
which the empty interval and the odd/even interval methods have been
suitably extended~\cite{DBA,MasAvrPLA,wenew,MetKlaPhysRep}. 
In particular, consider the
generalization to reversible coagulation, $A+A \rightleftharpoons
A$~\cite{DBA}.  For the ordinary diffusion-limited process this
generalization is accomplished by subtracting from the right hand side of
Eq.~(\ref{Eeqdifun}) a term proportional to 
$\partial E(x,t)/\partial x$ that describes the rate of decrease
of $E(x,t)$ due to the fact that particles
{\em adjacent} to either end of an empty interval may give birth (at
rate $v$) to a particle that moves into and therefore shrinks the
interval.  The same generalization is possible in the subdiffusive case.  
Assuming that the particles born at a rate $v$ move into an empty
interval subdiffusively, the evolution equation for 
$E(x,t)$ is given by
\begin{equation} 
\frac{\partial }{\partial t} E(x,t)=
~_{0}D_{t}^{1-\alpha }  \left\{ K_\alpha   \frac{\partial^2}{\partial x^2} E(x,t) - v
\frac{\partial}{\partial x} E(x,t)   \right\}.
\label{EreverGV}
\end{equation}
This equation can be solved by separation of
variables~\cite{MetKlaPhysRep,MetBarKlaPRL}.  The diffusive modes have
the same spatial dependence $\varphi_n(x)$ as in the diffusive case, but
their temporal evolution is no longer exponential.  Instead, it is
determined by the Mittag-Leffler function: 
$E(x,t)=\sum \varphi_n(x) E_\alpha (-\lambda_{n} t^\alpha)$,
where the $\lambda_n$ are the eigenvalues associated with the
$\varphi_n(x)$.
The equilibrium solution is
$E_{eq}
(x,t)=\exp(-vx/2K_\alpha)$, which in turn leads to
$c_{eq}=v/2K_\alpha$ and
$p_{eq}(x)=c_{eq}\exp(-c_{eq} x)$, as in the
case of ordinary diffusion~\cite{DBA,multi3}.  However, the
approach to equilibrium is algebraic rather than exponential and thus
qualitatively different,
$c(t)- c_{eq}\sim t^{-\alpha}$. 
An interesting attendant observation involves the dependence of these
results on the initial concentration $c(0)$.  In the diffusion-limited
case the time
constant $\tau$ in the exponential decay to equilibrium $\exp(-t/\tau)$
is a function of $c(0)$ for $c(0) < c_{eq}/2$ and changes to a
value independent of $c(0)$ for $c(0) > c_{eq}/2$~\cite{DBA}. 
In the subdiffusive
case it is the prefactor of $t^{-\alpha}$ that undergoes exactly the
same change.

We have considered the coagulation dynamics $A+A\rightarrow A$ and $A+A
\rightleftharpoons A$ and the annihilation dynamics $A+A \rightarrow 0$
for particles moving subdiffusively in one dimension.  This scenario
combines the ``anomalous kinetics" and ``anomalous diffusion" problems,
each of which leads to interesting dynamics separately and to even more
interesting dynamics in combination.  The fractional diffusion equation
plays a central role in our analysis and allows the exact calculation of
the density $c(t)$ and of the interparticle distribution function
$p(x,t)$ within this formulation.  Anomalous diffusion is characterized by
the exponent $\alpha$ 
introduced in Eq.~(\ref{meansquaredispl}), ordinary diffusion
corresponding to $\alpha=1$.  Deviations from ordinary diffusion lead to
a curious interplay.  On the
one hand, with decreasing $\alpha$ (and hence increasingly subdiffusive
motion) the decay of the particle density towards extinction or towards
equilibrium becomes increasingly slower and in this sense increasingly
different from law of mass action behavior.  On the other hand, the
spatial distribution of initially randomly distributed reactants
remains more Poissonian for all time as $\alpha$ decreases; indeed,
as $\alpha$ deviates from unity the relatively empty regions
around each particle
that are tantamount to (and indeed explain) anomalous kinetics in the
usual diffusion-limited case become more populated.

A number of generalizations and further questions that have been
considered in the context of diffusion-limited reactions but not for 
subdiffusion-limited reactions immediately come to
mind.  They include an analysis of different initial distributions, more
detailed consideration of reactions with external input (sources),
effective kinetic equations for the density, reactions in
statistically inhomogeneous
media, and reactions in the presence of external potentials.

This work has been supported in part by the Ministerio de Ciencia y
Tecnolog\'{\i}a (Spain) through Grant No. BFM2001-0718
and by the Engineering Research Program of
the Office of Basic Energy Sciences at the U. S. Department of Energy
under Grant No. DE-FG03-86ER13606.  SBY is also grateful to the DGES
(Spain) for a sabbatical grant (No.\ PR2000-0116).

\end{document}